\documentclass{moriondStyle}

\def\be{\begin{equation}}
\def\ee{\end{equation}}
\def\bea{\begin{eqnarray}}
\def\eea{\end{eqnarray}}

\begin{document}

\vspace*{4cm}
\title{Search for Double Beta Plus Decays with NuDoubt$^{++}$}

\author{C. GIRARD-CARILLO\,\footnote{\href{mailto:girardcarillo@uni-mainz.de}{girardcarillo@uni-mainz.de}}\\On behalf of the NuDoubt$^{++}$ collaboration\,\footnote{email: \href{mailto:nudoubt@lists.uni-mainz.de}{nudoubt@lists.uni-mainz.de}, website: \url{https://nudoubt.uni-mainz.de}}}

\address{Johannes Gutenberg University, Institute of Physics, Staudingerweg 7,\\
55128 Mainz, Germany}

\maketitle\abstracts{The NuDoubt$^{++}$ experiment proposes a novel detector concept to search for double beta plus decays using a hybrid opaque liquid scintillator.
  This design combines recent advances in scintillator technologies—namely, slow and opaque media, and optimized wavelength-shifting fibers coupled to silicon photomultipliers—to achieve improved particle identification and background rejection.
We present the physics motivations, detector design, first prototype layout, and expected sensitivities for both standard and exotic modes, using $^{78}$Kr as an initial target isotope.
With this approach, NuDoubt$^{++}$ aims to make the first observation of positron emitting Standard Model 2$\nu$ modes and improve current constraints on Beyond Standard Model 0$\nu$ processes.}

\section{Introduction: Motivation for Double Beta Plus Decay Searches}

Double beta decay is a second-order weak nuclear process, in which a nucleus emits two electrons (or two positrons) and two neutrinos~\cite{Goeppert-Mayer_1935}.
While the $2\nu\beta^-\beta^-$ mode is allowed in the Standard Model (SM), its neutrinoless counterpart ($0\nu\beta\beta$) would violate lepton number conservation and indicate that neutrinos are Majorana particles.
Experiments looking for this rare decay are key to understanding the nature of neutrinos and the origin of the Universe’s matter-antimatter asymmetry~\cite{Deppisch_2012}.

In contrast, the double beta plus decay modes ($2\nu\beta^+\beta^+$, $2\nu$EC$\beta^+$, and $2\nu$2EC, where EC stands for Electron Capture) are comparatively less explored.
These processes are suppressed due to smaller phase space, unfavorable values of the available energy of the decay $Q_{\beta\beta}$, and challenging detection signatures.
Nevertheless, they are critical to test nuclear matrix element calculations and probe lepton number violation through complementary isotopes~\cite{Hirsch_1994}. 

\section{Detector Concept: Hybrid and Opaque Scintillators}

To overcome the limitations of conventional detectors, NuDoubt$^{++}$ proposes a novel approach based on hybrid opaque scintillator technology.
Two key innovations are combined.

\subsection{Opaque Scintillators}

Traditional scintillators are made transparent to maximize light collection over long distances.
However, high loading with double beta isotopes reduces transparency.
Instead, NuDoubt$^{++}$ uses \emph{opaque scintillators} with short scattering lengths, confining the light near its creation point.
An embedded array of wavelength-shifting (WLS) fibers transports this light to SiPMs for readout.
The resulting 3D vertex localization allows particle identification (PID) and topological event reconstruction~\cite{LiquidO_2025}.

\subsection{Hybrid Slow Scintillators}

Slow scintillators delay the emission of scintillation light, allowing time separation from prompt Cherenkov radiation.
Since different particles like positron, electrons and $\gamma$'s emit different Cherenkov yields, the ratio of Cherenkov to scintillation light provides a powerful PID handle.
By preserving a high light yield, this technology improves energy resolution and event classification~\cite{Theia_2019}.

\section{The Detector Prototype Design}

\begin{figure}
\centerline{\includegraphics[width=0.55\linewidth]{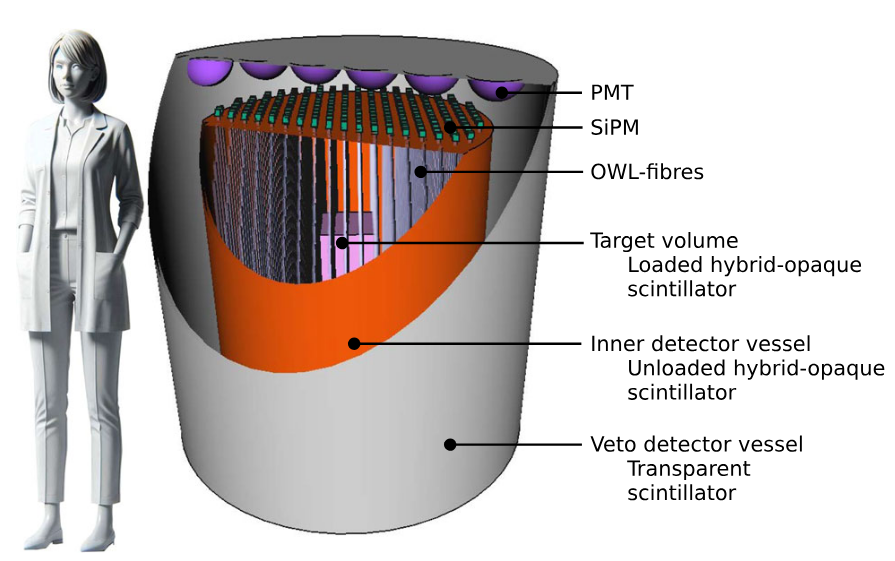}}
\caption[]{Basic detector layout: an outer veto with PMTs surrounds the inner detector filled with hybrid-opaque scintillator and OWL fibers, with a central 10~kg target volume at high pressure.}
\label{fig:design}
\end{figure}

The first NuDoubt$^{++}$ prototype presented in Figure~\ref{fig:design} is designed to validate the detection of double beta plus decays using a novel combination of hybrid and opaque scintillator techniques~\cite{Boehles2025}.
The core of the detector will consist of a high-pressure target volume filled with a hybrid-opaque liquid scintillator loaded with 50\% enriched krypton-78 gas.
The central vessel will operate at 5~bar and contains approximately 10~kg of active scintillating material, maximizing the number of double beta emitters.

To detect the light emitted from particle interactions within the scintillator, the detector is equipped with a dense grid of optimized wavelength-shifting (OWL) fibers~\cite{Kessler_2024,owls_2024}.
These fibers run along the axis of the cylindrical detector and are read out at both ends by silicon photomultipliers (SiPMs), allowing for three-dimensional reconstruction of the event topology using fiber position and timing, and precise measurement of the deposited energy.
The OWL fibers, specifically developed by the collaboration, provide enhanced light collection efficiency compared to commercial alternatives and are essential for achieving the desired spatial and energy resolutions.

Surrounding the central target is an optional active veto volume filled with a transparent liquid scintillator and instrumented with photomultiplier tubes (PMTs) on the top and bottom~\cite{DoubleChooz_2022_detector,AMOTech_2023}.
This layer serves to identify and reject external backgrounds such as cosmic muons or gamma radiation from surrounding materials.
The entire system is modular and designed for deployment in an underground facility such as LNGS (Laboratori Nazionali del Gran Sasso), where cosmic backgrounds are significantly suppressed.

This first-phase detector serves as a testbed for the NuDoubt$^{++}$ concept, with a clear focus on demonstrating the feasibility of detecting SM double beta plus decay modes and probing their BSM counterparts.

\section{Expected Sensitivities}

\begin{figure}
\begin{minipage}{0.49\linewidth}
\centerline{\includegraphics[width=0.95\linewidth]{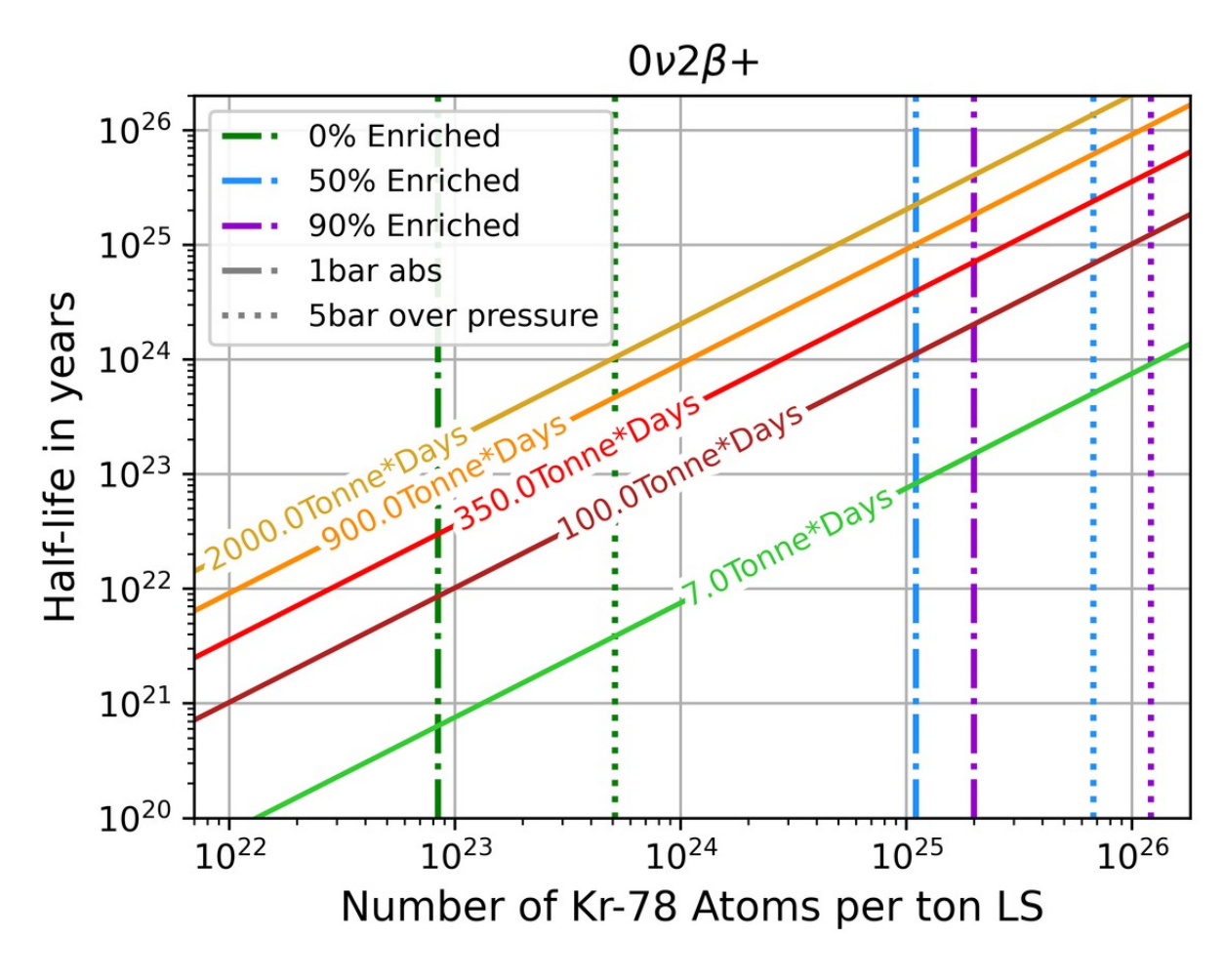}}
\end{minipage}
\hfill
\begin{minipage}{0.49\linewidth}
\centerline{\includegraphics[width=0.95\linewidth]{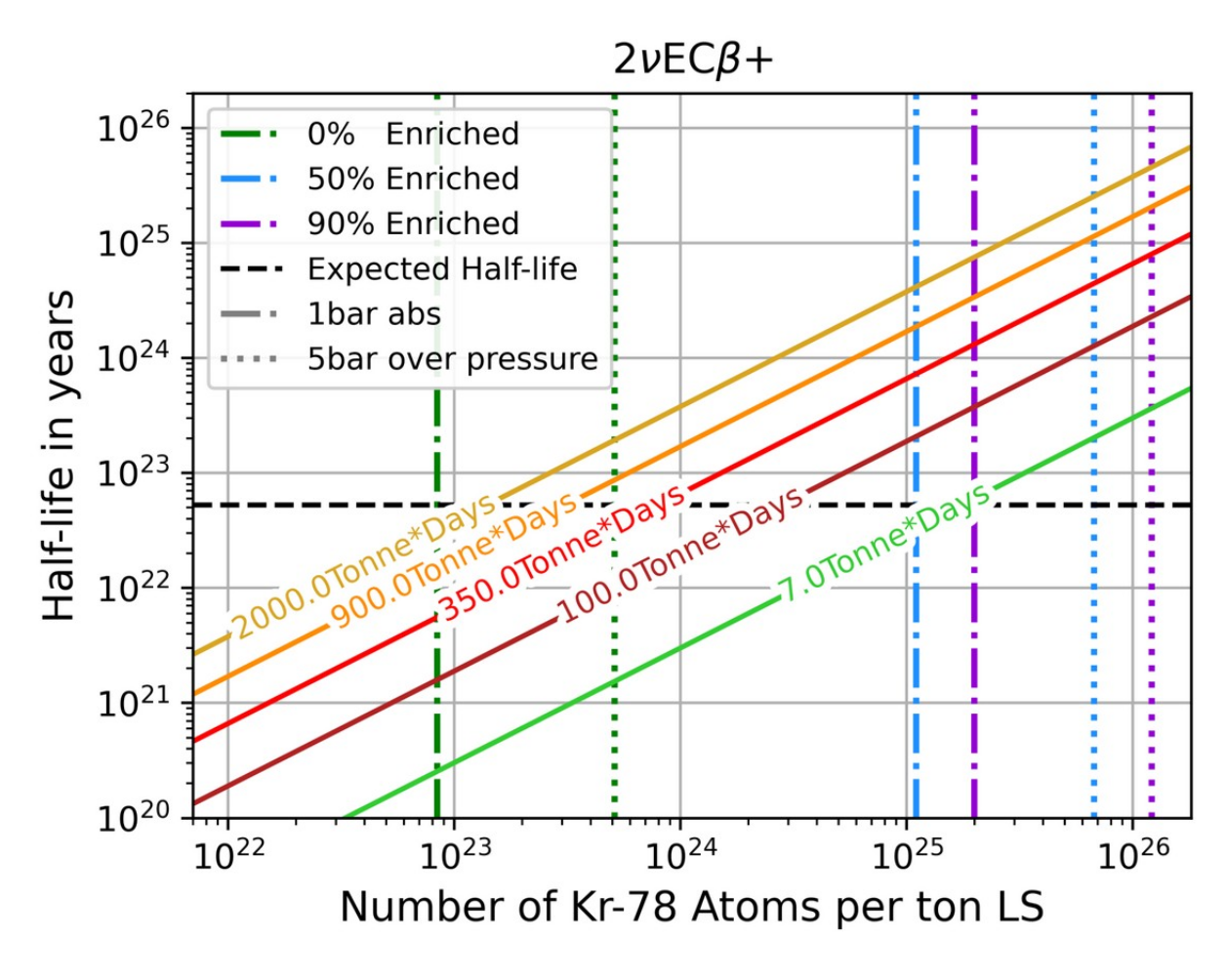}}
\end{minipage}
\caption[]{Left: Expected 90\% C.L. sensitivity on $0\nu\beta^+\beta^+$ half-life. Right: Discovery potential (5$\sigma$) for $2\nu$EC$\beta^+$ mode. Assumes Gran Sasso overburden. The prototype exposure is represented by the intersection between the green plain line ($7~$Tonne$\cdot$Days) and the blue vertical dashed line ($50~\%$ enrichment).}
\label{fig:sensitivity}
\end{figure}

The expected sensitivity for $2\nu$EC$\beta^+$ and the exclusion sensitivity for $0\nu2\beta^+$ are shown in Figure~\ref{fig:sensitivity}.
Sensitivity studies for the prototype assume one year of operation with a 10~kg target mass.

Using conservative assumptions on backgrounds and event selection, NuDoubt$^{++}$ is expected to achieve a 5$\sigma$ observation of the $2\nu$EC$\beta^+$ SM mode within this exposure.
For BSM processes, the experiment could reach a 90\% confidence level exclusion sensitivity for $0\nu\beta^+\beta^+$ half-lives up to $10^{24}$ years.
This would represent an improvement of nearly three orders of magnitude over current experimental limits.

\section{Current Status and Outlook}

Several key developments are currently underway within the NuDoubt$^{++}$ collaboration to validate and optimize the proposed detector concept.
A dedicated test bench is planned to be set up at the MAMI accelerator facility in Mainz, Germany, where a 25~cm $\times$ 25~cm prototype will be exposed to low energy electron beams and radioactive sources.
These tests are aimed at characterizing the detector response, assessing the performance of the OWL fibers, and verifying the particle identification and tracking capabilities enabled by the hybrid-opaque scintillator at these energies.

In parallel, ongoing laboratory efforts are focused on optimizing the coupling between OWL fibers and the scintillator medium.
Measurements of light yield and fiber compatibility are being conducted to optimize the optical efficiency of the system.
A dedicated test cell is also being developed to evaluate the feasibility of operating scintillator volumes under pressure while monitoring the loading and retention of enriched krypton gas.

Furthermore, the collaboration is working with the Max Planck Institute for Nuclear Physics (MPIK) in Heidelberg, Germany, to analyze the purity and isotopic composition of the krypton mixture using a precision proportional counter~\cite{Wink_1993}.
These measurements are essential for quantifying the active isotope content and minimizing contamination that could introduce unwanted backgrounds.

The combined outcome of these developments will orient the final design of the first demonstrator, guiding choices in scintillator formulation, isotope loading, and light collection architecture.
The NuDoubt$^{++}$ prototype is positioned as a pathfinder not only for detecting rare double beta plus decays, but also for pioneering the use of hybrid-opaque scintillators in future rare event searches, such as direct searches for light dark matter.

\section{Conclusion}

The NuDoubt$^{++}$ experiment introduces an innovative approach to the search for double beta plus decays, combining hybrid slow and opaque scintillator technologies with advanced light collection using OWL fibers and SiPMs.
A prototype is under active development, with promising sensitivities to both Standard Model and neutrinoless modes.
These efforts are summarized in our first publication, which lays the groundwork for future experimental exploration in this direction~\cite{Boehles2025}.

\section*{Acknowledgments}

This work has been supported by the Cluster of Excellence ``Precision Physics, Fundamental Interactions, and Structure of Matter'' (PRISMA$^+$ EXC 2118/1) funded by the German Research Foundation (DFG) within the German Excellence Strategy (Project ID 390831469).
We are especially thankful for the support of the PRISMA Detector Laboratory.

\section*{References}
\bibliography{bibliography}

\begin{thebibliography}{10}
\providecommand{\url}[1]{#1}
\csname url@samestyle\endcsname
\providecommand{\newblock}{\relax}
\providecommand{\bibinfo}[2]{#2}
\providecommand{\BIBentrySTDinterwordspacing}{\spaceskip=0pt\relax}
\providecommand{\BIBentryALTinterwordstretchfactor}{4}
\providecommand{\BIBentryALTinterwordspacing}{\spaceskip=\fontdimen2\font plus
\BIBentryALTinterwordstretchfactor\fontdimen3\font minus
  \fontdimen4\font\relax}
\providecommand{\BIBforeignlanguage}[2]{{%
\expandafter\ifx\csname l@#1\endcsname\relax
\typeout{** WARNING: IEEEtran.bst: No hyphenation pattern has been}%
\typeout{** loaded for the language `#1'. Using the pattern for}%
\typeout{** the default language instead.}%
\else
\language=\csname l@#1\endcsname
\fi
#2}}
\providecommand{\BIBdecl}{\relax}
\BIBdecl

\bibitem{Goeppert-Mayer_1935}
\BIBentryALTinterwordspacing
M.~Goeppert-Mayer, ``Double beta-dis\-integration,'' \emph{Phys. Rev.},
  vol.~48, pp. 512--516, 9 1935.  \url{https://doi.org/10.1103/PhysRev.48.512}
\BIBentrySTDinterwordspacing

\bibitem{Deppisch_2012}
\BIBentryALTinterwordspacing
F.~Deppisch, M.~Hirsch, and H.~Päs, ``Neutrinoless double-beta decay and
  physics beyond the standard model,'' \emph{J. Phys. G}, vol.~39, no.~12, p.
  124007, Nov. 2012.  \url{http://dx.doi.org/10.1088/0954-3899/39/12/124007}
\BIBentrySTDinterwordspacing

\bibitem{Hirsch_1994}
\BIBentryALTinterwordspacing
M.~Hirsch, K.~Muto, T.~Oda, and H.~V. Klapdor-Kleingrothaus, ``{Nuclear
  structure calculation of $\beta$+$\beta$+, $\beta$+/EC and EC/EC decay matrix
  elements},'' \emph{Z. Phys. A}, vol. 347, no.~3, pp. 151--160, 9 1994.
  \url{https://doi.org/10.1007/BF01292371}
\BIBentrySTDinterwordspacing

\bibitem{LiquidO_2025}
\BIBentryALTinterwordspacing
J.~Apilluelo \emph{et~al.}, ``{The Stochastic Light Confinement of LiquidO},''
  \emph{arXiv}, 2025.  \url{https://arxiv.org/abs/2503.02541}
\BIBentrySTDinterwordspacing

\bibitem{Theia_2019}
\BIBentryALTinterwordspacing
M.~Askins \emph{et~al.}, ``{THEIA: an advanced optical neutrino detector},''
  \emph{Eur. Phys. J. C}, vol.~80, no.~5, p. 416, 2020.
  \url{http://doi.org/10.1140/epjc/s10052-020-7977-8}
\BIBentrySTDinterwordspacing

\bibitem{Boehles2025}
\BIBentryALTinterwordspacing
M.~Böhles \emph{et~al.}, ``Combining hybrid and opaque scintillator techniques
  in the search for double beta plus decays,'' \emph{Eur. Phys. J. C}, vol.~85,
  no.~2, p. 121, 2025.  \url{https://doi.org/10.1140/epjc/s10052-025-13847-1}
\BIBentrySTDinterwordspacing

\bibitem{Kessler_2024}
\BIBentryALTinterwordspacing
B.~Ke{\ss}ler, ``Development of wavelength-shifting fibers with high photon
  capture-rate,'' Master's thesis, Johannes Gutenberg University Mainz, 2024.
  \url{https://nudoubt.uni-mainz.de/theses/2024_Master_Kessler_Bastian.pdf}
\BIBentrySTDinterwordspacing

\bibitem{owls_2024}
B.~Keßler, J.~Rack-Helleis, and S.~Böser, ``Cylindrical light guides with
  near-surface emission centers,'' no. DE 10 2023 135 496.5 Patent Application,
  2024, {German Patent and Trade Mark Office}.

\bibitem{DoubleChooz_2022_detector}
\BIBentryALTinterwordspacing
H.~de~Kerret \emph{et~al.}, ``{The Double Chooz antineutrino detectors},''
  \emph{Eur. Phys. J. C}, vol.~82, no.~9, p. 804, 2022.
  \url{http://doi.org/10.1140/epjc/s10052-022-10726-x}
\BIBentrySTDinterwordspacing

\bibitem{AMOTech_2023}
\BIBentryALTinterwordspacing
A.~Cabrera, ``{CLOUD / AntiMatter-OTech: A New Generation of Neutrino Science
  at Chooz},'' 2023.  \url{https://doi.org/10.5281/zenodo.10049846}
\BIBentrySTDinterwordspacing

\bibitem{Wink_1993}
\BIBentryALTinterwordspacing
R.~Wink \emph{et~al.}, ``{The miniaturized proportional counter HD-2(Fe)/(Si)
  for the GALLEX solar neutrino experiment},'' \emph{Nucl. Instrum. Meth. A},
  vol. 329, no.~3, pp. 541--550, 1993.
  \url{https://www.sciencedirect.com/science/article/pii/016890029391289Y}
\BIBentrySTDinterwordspacing

\end{thebibliography}

\end{document}